# Preprint Big City 3D Visual Analysis

Zhihan Lv[†1] and Xiaoming Li[1,2,3] and Baoyun Zhang[4] and Weixi Wang[2,3] and Shengzhong Feng[1] and Jinxing Hu[1]

[1]Shenzhen Institutes of Advanced Technology(SIAT), Chinese Academy of Science, China
[2]Shenzhen Research Center of Digital City Engineering, Shenzhen, China
[3]Key Laboratory of Urban Land Resources Monitoring and Simulation, Ministry of Land and Resources, Shenzhen, China
[4]Jining Institutes of Advanced Technology(JIAT), Chinese Academy of Science, China

**Abstract**
*A big city visual analysis platform based on Web Virtual Reality Geographical Information System (WEBVRGIS) is presented. Extensive model editing functions and spatial analysis functions are available, including terrain analysis, spatial analysis, sunlight analysis, traffic analysis, population analysis and community analysis.*

Categories and Subject Descriptors (according to ACM CCS): Computer Graphics [1.3.7]: Three-Dimensional Graphics and Realism—Virtual reality

## 1. Introduction

Virtual Reality Geographical Information System (VRGIS) has been widely used for diverse applications such as landscape planning, resource management, urban modeling and etc. It becomes ever important in the era of big data [big08] [Bri12], and functions as a fundamental infrastructure of smart cities. Coming along with the high demands is higher challenge, mostly regarding to the capacity of managing multi-source geospatial data sets, integrating into a uniformed structure, performing advanced analysis and simulation, effectively visualizing large volumes of simulated and real-time data, and finally, sharing information by efficient means [BZ11]. In view of this situation, we consider WebVRGIS platform as a solution to meet these requirements, furthermore, we find that many issues can be overcome by enhanced virtual reality technology. This work presents a mature system developed in such line, which is loaded with 3D Shenzhen data as a demonstration.

Nowadays, there is an increasing interest in creating Virtual Reality Geographical Information System (VRGIS), which can obtain the landscape geospatial data dynamically, as well as perform rich visual 3D analysis, calculations, managements based on Geographical Information System (GIS) data. The web version of VRGIS is so-called WebVRGIS, which is also widely known as virtual geographical environments. GIS data has several characteristics, such as large scale, diverse predictable and real-time, which falls in the range of definition of Big Data [Bri12]. Besides, to improve the accuracy of modeling, the city planning has an increasingly high demand for the realistic display of VR system, however this will inevitably lead to the growth of the volume of data. Virtual scene from a single building to the city scale is also resulting in the increased amount of data. In addition, the concern of usability of WebVRGIS has attracted the attentions to a new challenge, which is fusing all kinds of city information big data by WebVRGIS platform, and thus exploiting the data effectively. Beside spatial data integration, new user interfaces for geo-databases is also expected [BZ11]. Therefore, the management and development of city big data using virtual reality technology is a promising and inspiring approach.

† lvzhihan@gmail.com

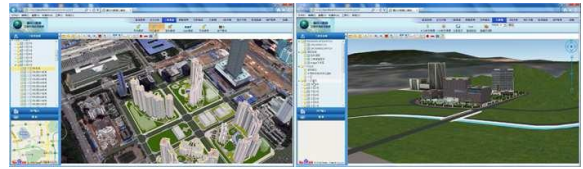

**Figure 1:** *The UI of the proposed system.*

## 2. Data

We utilize 3D Shenzhen as a convincing case to present WebVRGIS [Le13], as shown in figure 1. Shenzhen is a thirty-years new city, however, it has the highest population density in China, which reaches 7785 people per square kilometer (2013). It causes some embarrassments to the city information management. The presented platform can assistant the social service agencies to make full use of Virtual Reality and database technology to improve the comprehensive level of the city's service management, it can achieve shares of information resources of all departments and the dynamic tracking for the population and companies, geospatial information.

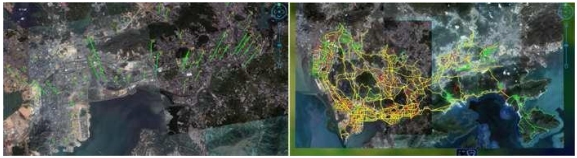

**Figure 2:** *Left: Passenger flow of various stations; Right: Real-time traffic visualization*

## 3. System Design

**Graphic User Interface** WebVR render engine [LYH*11] and spatial index technology is adopted to realize high-performance mapping of massive model data, realize special effects of rich scenes and smooth roaming, and truly simulate 3D urban space environment. Extensive model editing functions and spatial analysis functions are available, including terrain analysis, spatial analysis, sunlight analysis, traffic analysis, population analysis and community analysis. Users can realize dynamic loading of various kinds of spatial data, and change plane GIS data into more intuitive 3D symbols, as shown in figure 1.

On the website part, the top view-based position observation is conducted through a 2D bird's-eye map on the left. The tool bar on the top of the earth browser contains 3D GIS analysis function and urban data management, analysis and visualization. The tree-view management control on the upper left contains various layers in the earth browser. By clicking the tree view management control on the left, it is possible to display the 3D model of each building and each room respectively.

**Big data handling and analysis** Virtual community [LCZ*12] [ZLZ*09] based big data management service is used to supply big data access service to various city information management systems and to manage the big data system. Various city information management systems can be enabled to obtain data and information from the big data platform in the form of service through access control over the data aggregation processing platform. Analysis function can have a comprehensive observation to the 3D model of a community and analyze and display various data, including population age composition, education background composition, etc.

It counts the population distribution in various areas of the city and can manage the urban traffic and real-time road condition, as well as forecast the passenger flow, as shown in figure 2.

This platform can manage the urban traffic and real-time road condition. It can visualize the road condition in the form of line and plane. In addition to real-time data, it can also load the historical data. This platform also manages the data in bus and taxi stations and subway stations in Shenzhen and forecasts the passenger flow, as shown in figure. Through clicking the button of âĂIJdeformation along metro lines âĂİ (100m or 50m), the system will analyze the deformation along metro lines (100m or 50m), and show the result in a way of spatial cylinder or spatial point to represent the trend of deformation. The cylinder or icon above ground shows lifting deformation, and the cylinder or icon below ground shows sinking deformation; in case of larger height difference, it means larger deformation.

## 4. Implementation

The platform is developed by C++, OpenGL for rendering and HTML, javascript, C# for website distribution. The city bigdata contains the land and ocean [SLG*14] [LS14] [LSLF15], the above-ground and underground, outdoor and indoor, building and people, real-time and history as well as the forecast [LLZ*15]. Spatiotemporal database model and visualization has been considered in the design of the system [ZWZA12].

## 5. Conclusion

The 3D Shenzhen case proves 3D city visualization and analysis platform is a useful tool for the social service agencies and citizens for browsing and analyzing city big data directly, and is agreed upon as being both immediately useful and generally extensible for future applications. It also demonstrates that WebVRGIS engine is a state-of-the-art practical research. Some novel interaction approaches are considered to integrate in our future work [Lv13] [LFLF14] [LFFL15].

### Acknowledgments

The authors are thankful to the National Natural Science Fund for the Youth of China (41301439) and Electricity 863 project(SS2015AA050201).

### References

[big08] Big data specials. *Nature 455*, 7209 (Sept. 2008). 1

[Bri12] BRIGGS F.: Large data - great opportunities. Presented at IDF2012, Beijing. 1

[BZ11] BREUNIG M., ZLATANOVA S.: Review: 3d geo-database research: Retrospective and future directions. *Comput. Geosci. 37*, 7 (July 2011), 791–803. 1

[LCZ*12] LV Z., CHEN G., ZHONG C., HAN Y., QI Y. Y.: A framework for multi-dimensional webgis based interactive online virtual community. *Advanced Science Letters 7*, 1 (2012), 215–219. 2

[Le13] LV Z., ET.AL.: Webvrgis: A p2p network engine for vr data and gis analysis. In *ICONIP 2013*. 2013. 1

[LFFL15] LV Z., FENG L., FENG S., LI H.: Extending touchless interaction on vision based wearable device. In *Virtual Reality (VR), 2015 iEEE* (2015), IEEE. 2

[LFLF14] LV Z., FENG L., LI H., FENG S.: Hand-free motion interaction on google glass. In *SIGGRAPH Asia 2014 Mobile Graphics and Interactive Applications* (2014), ACM. 2

[LLZ*15] LI X., LV Z., ZHANG B., WANG W., FENG S., HU J.: Webvrgis based city bigdata 3d visualization and analysis. In *Visualization Symposium (PacificVis), 2015 IEEE Pacific* (2015), IEEE. 2

[LS14] LV Z., SU T.: 3d seabed modeling and visualization on ubiquitous context. In *SIGGRAPH Asia 2014 Posters* (2014), ACM, p. 33. 2

[LSLF15] LV Z., SU T., LI X., FENG S.: 3d visual analysis of seabed on smartphone. In *Visualization Symposium (PacificVis), 2015 IEEE Pacific* (2015), IEEE. 2

[Lv13] LV Z.: Wearable smartphone: Wearable hybrid framework for hand and foot gesture interaction on smartphone. In *2013 IEEE International Conference on Computer Vision Workshops* (2013), IEEE, pp. 436–443. 2

[LYH*11] LV Z., YIN T., HAN Y., CHEN Y., CHEN G.: Webvr–web virtual reality engine based on p2p network. *Journal of Networks 6*, 7 (2011). 2

[SLG*14] SU T., LV Z., GAO S., LI X., LV H.: 3d seabed: 3d modeling and visualization platform for the seabed. In *Multimedia and Expo Workshops (ICMEW), 2014 IEEE International Conference on* (2014), IEEE, pp. 1–6. 2

[ZLZ*09] ZHANG M., LV Z., ZHANG X., CHEN G., ZHANG K.: Research and application of the 3d virtual community based on webvr and ria. *Computer and Information Science 2*, 1 (2009), P84. 2

[ZWZA12] ZHONG C., WANG T., ZENG W., ARISONA S. M.: Spatiotemporal visualisation: A survey and outlook. In *Digital Urban Modeling and Simulation*. Springer, 2012, pp. 299–317. 2